\documentclass[10pt,letterpaper]{article}

\usepackage{cogsci}

\cogscifinalcopy 

\usepackage{pslatex}
\usepackage{apacite}
\usepackage{float} 

\usepackage{subcaption}
\usepackage{graphicx}
\graphicspath{{figures}}
\usepackage{makecell}
\usepackage{bm}
\usepackage{multirow}
\usepackage{footmisc}
\usepackage[table]{xcolor}
\usepackage{hhline}
\usepackage{wrapfig}

\newlength{\Oldarrayrulewidth}
\newcommand{\Cline}[2]{%
  \noalign{\global\setlength{\Oldarrayrulewidth}{\arrayrulewidth}}%
  \noalign{\global\setlength{\arrayrulewidth}{#1}}\cline{#2}%
  \noalign{\global\setlength{\arrayrulewidth}{\Oldarrayrulewidth}}}

\title{The Power of Nudging: Exploring Three Interventions for \\Metacognitive Skills Instruction across Intelligent Tutoring Systems}

\author{{ \large \bf Mark Abdelshiheed, John Wesley Hostetter, Preya Shabrina, Tiffany Barnes, and Min Chi} \\
  Department of Computer Science\\
  North Carolina State University\\
  Raleigh, NC 27695 \\
  \{{mnabdels,\, jwhostet,\, pshabri,\, tmbarnes,\, mchi\}}@ncsu.edu}

\begin{document}

\maketitle

\begin{abstract}

Deductive domains are typical of many cognitive skills in that no single  problem-solving strategy is always optimal for solving all problems. It was shown that students who know \emph{how} and \emph{when} to use each strategy $(StrTime)$ outperformed those who know neither and stick to the default strategy $(Default)$. In this work, students were trained on a logic tutor that supports a \emph{default} forward-chaining and a backward-chaining (BC) strategy, then a probability tutor that only supports BC. We  investigated three types of interventions on teaching the $Default$ students \emph{how} and \emph{when} to use which strategy on the logic tutor: $Example$, $Nudge$ and $Presented$. Meanwhile, $StrTime$ students received no interventions. Overall, our results show that $Nudge$ outperformed their $Default$  peers and caught up with $StrTime$ on both tutors.


\end{abstract}

\section{Introduction}

Deductive task domains are those in which a solution requires an argument, proof, or derivation; each step is the outcome of applying a  domain principle, operator, or rule. Deductive domains such as geometry, logic and probability are standard components of STEM fields.  Two common problem-solving strategies in such domains are forward-chaining (FC) and backward-chaining (BC) \cite{russell2020artificial}. In FC, the reasoning proceeds from the given propositions toward the target goal, whereas BC is goal-driven in that it works backward from a goal state to a given state. Early studies show that experts often use a mixture of FC and BC strategies, and more importantly, they often use past experience, heuristics, and many other kinds of knowledge to determine their strategies \cite{priest1992newINSTRUCTION}. Our prior work showed that  students who know \emph{which} problem-solving strategies to use \emph{when}, referred to as $StrTime$, consistently learn across different deductive domains, as they possess the necessary \emph{metacognitive skills}, unlike their peers who follow the default strategy, known as $Default$ (Abdelshiheed et al., 2020).

It has been believed that metacognitive skills are essential for academic achievements \cite{de2018longSWITCH-INSTRUCTION, erskine2010metacognitiveInstruction, zimmerman1990Metacognition}, and teaching such skills impacts learning outcomes \cite{zepeda2015INSTRUCTION, chi2010backward2metacogStrategyINSTRUCTION} as well as strategy use \cite{lee2008koreanStrategyAwareness, chambres2002metacognitionStrategySelection,roberts1993metacogDefinitionStrategySelection2}. STEM domains often demand the use of various problem-solving strategies, and some prior research has categorized knowing \emph{how} and \emph{when} to use each strategy as two metacognitive skills \cite{winne2014switchMetacognitiveSWITCH,cardelle1992effectsMETACOGNITIVE}, referred to as strategy- and time-awareness, respectively. 

Prior work has shown the positive impact of strategy awareness on preparing students for future learning \cite{belenky2012strategyAwarenessPFL,abdelshiheed2021preparing} and time awareness on planning skills \cite{winne2014switchMetacognitiveSWITCH,fazio2016timeAwareness}. Thus various attempts were made to teach students the two metacognitive skills, such as teaching the strategy by example \cite{likourezos2017WE,glogger2015WE}, prompting nudges to use the strategy \cite{richey2015promptsWE,belenky2009metacogTransfer} and explicitly presenting it \cite{fellman2020Presented,sporer2009Presented}.

Our work directly compares three types of interventions on teaching $Default$ students \emph{how} and \emph{when} to use which strategy on the logic tutor in this ascending order of instructional support: $Example$, $Nudge$ and $Presented$. All interventions provided BC worked examples. The main difference is that $Nudge$ prompted students to switch to BC in problems proper to do so, while for $Presented$, those problems were presented in BC by default. Our primary research question is: Which of the three types of interventions would make $Default$ students catch up with their $StrTime$ peers?

Our study involved two intelligent tutoring systems (ITSs) \cite{vanlehn2006behavior}: logic and probability. Students were first assigned to a logic tutor that supports FC and BC strategies, with FC being the default, then to a probability tutor six weeks later that supports BC only. During the logic instruction, $Default$ students were split into four conditions: three  intervention groups ---$Example$, $Nudge$ and $Presented$--- and a $Control$ group without any intervention. On the other hand, we believe that $StrTime$ students already have the two metacognitive skills and thus are considered the gold standard and received no intervention. All students went through the same probability tutor and were asked to decide whether they wanted to solve the following problem on their own (problem-solving $(PS)$), the tutor to present it as a worked example $(WE)$, or to solve it collaboratively with the tutor in the form of a faded worked example $(FWE)$. Overall, our results show that $Nudge$ students outperformed their other $Default$ peers and caught up with $StrTime$ on both tutors. Additionally, $Nudge$’s strategy behavior on the logic tutor was similar to $StrTime$ as both knew how and when to use BC. Surprisingly, $Nudge$ chose significantly more $PS$ on the probability tutor, and $StrTime$ chose significantly less $FWE$.

\section{Related Work}

\subsection{Teaching by Example, Nudging and Presenting}

Substantial work has explored many approaches for teaching strategies and highlighted their tradeoffs. We focus on the possible combinations of three approaches: teaching a strategy by example \cite{likourezos2017WE,glogger2015WE}, prompting nudges to use a strategy \cite{richey2015promptsWE,zepeda2015INSTRUCTION,belenky2009metacogTransfer} and directly presenting it \cite{fellman2020Presented,sporer2009Presented, chi2007pyrenees2,schwartz2004Presented}.

\citeA{glogger2015WE} found that students receiving worked examples of journal extracts reviews outperformed their peers, who had to come up with the reviews, on post-test performance. However, \citeA{likourezos2017WE} reported no significant difference between students who received fully-guided worked examples, partially-guided ones and unguided assistance on post-test geometry tasks.

\citeA{zepeda2015INSTRUCTION} showed that students who received tutoring nudges and worked examples performed better on a physics test and a novel self-guided activity than their peers who received no such instruction. Conversely, \citeA{richey2015promptsWE} found no significant difference between students who were instructed to study the worked examples and their peers, who received the same examples with tutoring nudges, on near, intermediate and far transfer electric circuit tasks.

\citeA{sporer2009Presented} found that students who were explicitly instructed on comprehensive reading strategies surpassed their peers, who were taught by the instructors' text interactions, on a transfer task and follow-up test.  On the other hand, \citeA{fellman2020Presented} found no significant difference between students who were presented explicit strategy instruction to practice the single-digit n-back task and their peers who practiced without such instruction, as both groups showed emerging transfer to untrained variants of the same task.

\subsection{Metacognitive Skills Instruction}

Metacognitive skills regulate one's awareness and control of their  cognition \cite{chambres2002metacognitionStrategySelection,roberts1993metacogDefinitionStrategySelection2}. Many studies have demonstrated the significance of metacognitive skills instruction on academic performance \cite{de2018longSWITCH-INSTRUCTION,erskine2010metacognitiveInstruction}, learning outcomes \cite{zepeda2015INSTRUCTION,chi2010backward2metacogStrategyINSTRUCTION,chi2008eliminatingGap} and regulating strategy use \cite{schraw2015metacognitiveInstruction}.

\citeA{schraw2015metacognitiveInstruction} argue that metacognitive skill instruction involves feeling what is known and not known about a task, as this allows learners to gather information efficiently, adapt to changes in task requirements, and develop strategies to overcome the task. They state that such instruction should further compare strategies according to their feasibility and familiarity from the learner's perspective.

\citeA{belenky2009metacogTransfer} showed that students who were prompted with metacognitive nudges, which reflect on the current problem-solving processes, outperformed their peers who received problem-focused nudges, which focus on the current goal, on a permutation transfer task. \citeA{chi2010backward2metacogStrategyINSTRUCTION} found that teaching students principle-emphasis skills closed the gap between high and low learners, not only in the domain where they were taught (probability) but also in a second domain where they were not taught (physics).

\subsection{Strategy- and Time-Awareness}

Strategy- and time-awareness have been regarded as metacognitive skills as they respectively address \emph{how} and \emph{when} to use a problem-solving strategy \cite{de2018longSWITCH-INSTRUCTION,winne2014switchMetacognitiveSWITCH,lee2008koreanStrategyAwareness,cardelle1992effectsMETACOGNITIVE}. Researchers have emphasized the role of strategy awareness in learning a foreign language \cite{teng2020strategyAwareness,lee2008koreanStrategyAwareness} and preparation for future learning \cite{abdelshiheed2021preparing,belenky2012strategyAwarenessPFL, chamot1998strategyAwareness}, and the impact of time awareness on planning skills and academic performance \cite{de2018longSWITCH-INSTRUCTION, fazio2016timeAwareness, winne2014switchMetacognitiveSWITCH}.

\citeA{lee2008koreanStrategyAwareness} studied the role of strategy awareness in teaching English to Korean students; specifically, students aware of various learning strategies employed these strategies more frequently than their peers. In \citeA{abdelshiheed2021preparing}, we found that students who knew two problem-solving strategies were the best learners in two independent domains. \citeA{belenky2012strategyAwarenessPFL} showed that students who had a higher aim to master presented materials and strategies outperformed their peers on a transfer task.

In \citeA{fazio2016timeAwareness}, students who knew when to use each strategy to pick the largest fraction magnitude had high mathematical proficiency. Their peers who did not know when to apply each strategy failed to choose the correct alternative when offered choices. \citeA{de2018longSWITCH-INSTRUCTION} showed that students who knew when and why to use a given strategy exhibit long-term metacognitive knowledge that improves their academic performance. \citeauthor{de2018longSWITCH-INSTRUCTION} emphasized that knowing \emph{when} and \emph{why} has the same importance as knowing \emph{how} when it comes to strategy choice in multi-strategy domains.

To sum up, much of the prior work has highlighted the importance of metacognitive skill instruction and teaching strategy- and time-awareness. Many approaches for teaching strategies have been investigated, such as teaching by example, prompting nudges, and direct presentation. However, as far as we know, no agreement has been found on the most effective combination of these approaches, and no work has compared these approaches in intelligent tutoring systems. This work compares three ways to teach a backward-chaining (BC) strategy on two intelligent tutoring systems: logic and probability. First, by examples alone $(Example)$, then by examples and nudges to switch to BC $(Nudge)$, and finally, by examples and directly presenting BC $(Presented)$.

\section{Methods}

\begingroup
\renewcommand{\arraystretch}{1.7}
\begin{table}[ht!]
\scriptsize
\begin{center} 
\caption{Tutors' Assignment and Completion Counts} 
\label{completionRate} 
\begin{tabular}{lcc|cc}
\Xhline{4\arrayrulewidth}
& \multicolumn{2}{c}{Logic} & \multicolumn{2}{c}{Probability}\\
\cline{2-5}
& \makecell[t]{Assigned} & \makecell[t]{Completed} & \makecell[t]{Assigned}   & \makecell[t]{Completed} 
  \\
\hline
$Control$ & $23$  & $21$ &  $19$ &  $\mathbf{17}$  \\
$Example$ & $23$ & $20$ &  $20$ &  $\mathbf{19}$ \\
$Nudge$ & $22$  & $21$ &  $21$ &  $\mathbf{20}$  \\
$Presented$ & $20$ & $17$ &  $16$ &  $\mathbf{15}$ \\
$StrTime$ & $49$  & $45$ &  $41$ &  $\mathbf{40}$  \\
\hline
& \multicolumn{2}{c|}{\makecell[t]{$\chi^2 (4,\, N=261) = 0.09,\,$ \\ $p=.99$}} & \multicolumn{2}{c}{\makecell[t]{$\chi^2 (4,\, N=228) = 0.05,\,$ \\ $p=.99$}} \\

\Xhline{4\arrayrulewidth}
\end{tabular} 
\end{center} 
 {\centering Only students who completed Logic were assigned to Probability.\par}
\end{table}
\endgroup

\nocite{abdelshiheed2020metacognition}
\nocite{abdelshiheed2022assessing}

\subsection{Participants}
They are Computer Science undergraduates at North Carolina State University. Each tutor is a class assignment whose completion is required for full credit, and students are told that grades are based on effort, not performance. The main challenge in this work is that the student's metacognitive label ---$Default$ or $StrTime$--- can be calculated only at the end of logic training, but the label is needed at its beginning to determine the intervention possibility. Specifically, $StrTime$ students frequently follow the desired behavior of switching \emph{early} (within the first $30$ actions) to $BC$, while $Default$ students make no switches and stick to $FC$ (Abdelshiheed et al., 2022, 2020). Such switch behaviors are recorded at the end of the logic training, and hence, can not be calculated before training. Therefore, as per \citeA{abdelshiheed2021preparing}, we utilize the random forest classifier (RFC) that, based on pre-test performance, predicts the metacognitive label before training on logic and was previously shown to be $96\%$ accurate.

Among $230$ students assigned to the logic tutor, $137$ were classified by the RFC into $88$ $Default$ and $49$ $StrTime$\footnote{The remaining students were excluded from further analyses, as their label is irrelevant to this work}. $Default$ students were randomly split into four conditions: a control ---$Control$--- and three experimental ---$Example$, $Nudge$ and $Presented$. Table \ref{completionRate} shows the assigned and completed counts on both tutors for $Default$ (top four rows) and $StrTime$ (fifth row). The last column is for students who finished both tutors since we excluded dropout logic students from the probability assignment. Hence, only the last column students were included in our analyses resulting in $17$ $Control$, $19$ $Example$, $20$ $Nudge$, $15$ $Presented$ and $40$ $StrTime$. As shown in Table \ref{completionRate}, a chi-square test found no significant difference between the groups' completion rates on both tutors. The RFC was $97\%$ accurate in classifying students who received no interventions ---$Control$ and $StrTime$.

\subsection{Two Tutors and Our  Interventions}

\begin{figure}[ht!]
     \centering
     \begin{subfigure}[t]{0.23\textwidth}
         \centering
         \includegraphics[width=\textwidth,height =4cm]{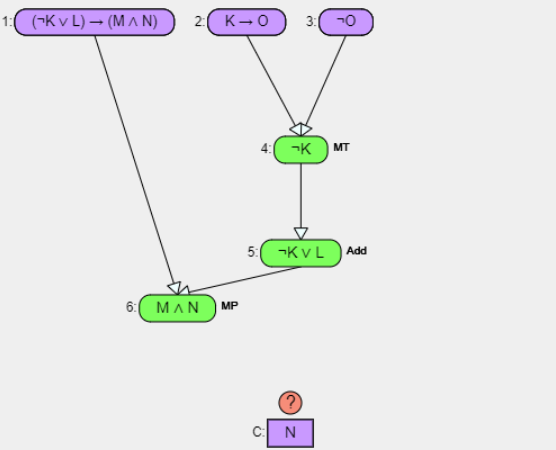}
         \caption{Forward Chaining (FC)}
         \label{fig:direct}
     \end{subfigure}
     \hfill
     \begin{subfigure}[t]{0.24\textwidth}
         \centering
         \includegraphics[width=\textwidth,height=4cm]{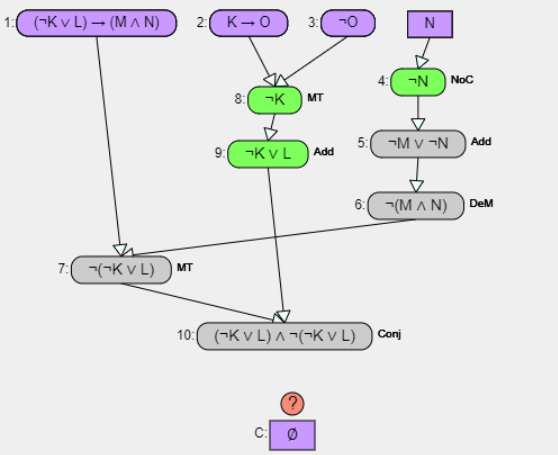}
         \caption{Backward Chaining (BC)}
         \label{fig:indirect}
     \end{subfigure}
\caption{Logic Tutor Problem-Solving Strategies}
\label{DT}
\end{figure}

\begin{figure}[ht!]
\begin{center}
\includegraphics[width=6cm, height = 3.5cm]{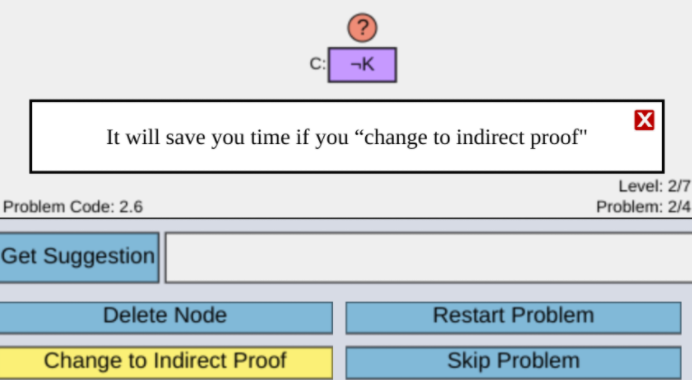}
\end{center}
\caption{Prompted Strategy Switch Nudge} 
\label{fig:prompt}
\end{figure}

\subsubsection{Logic Tutor and Our Interventions}
\begin{figure}[ht!]
\begin{center}
\includegraphics[width=0.38\textwidth]{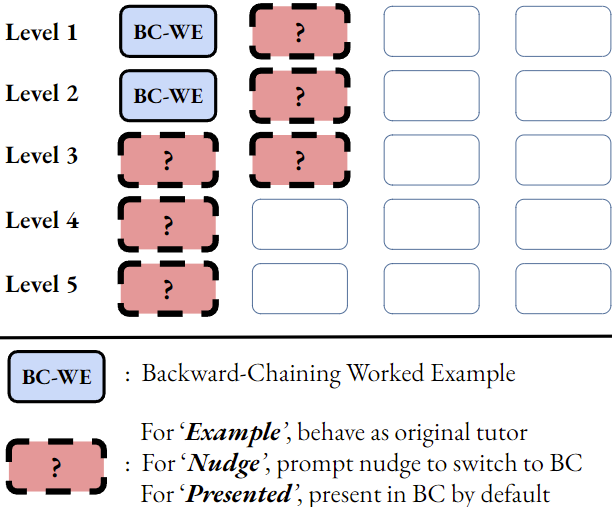}
\end{center}
\caption{Training on the Adjusted Logic Tutor} 
\label{fig:modified_logic}
\end{figure}

The logic tutor teaches propositional logic proofs by applying valid inference rules such as Modus Ponens and Constructive Dilemma. It consists of five ordered levels with an \emph{incremental degree of difficulty}, and each level consists of four problems. A student can solve any problem by either a \textbf{FC} or \textbf{BC} strategy. Figure \ref{fig:direct} shows that in \emph{FC}, one must derive the conclusion at the bottom from givens at the top, while Figure \ref{fig:indirect} shows that in \emph{BC}, students derive a contradiction from givens and the \emph{negation} of the conclusion. Problems are presented by \emph{default} in FC with the ability to switch to BC by clicking the yellow button in Figure \ref{fig:prompt}. The logic tutor was adjusted, as shown in Figure \ref{fig:modified_logic}, to accommodate the following interventions for $Default$ students:
\begin{itemize}
    \setlength\itemsep{0.01em} 
    \item \textit{No Intervention}: students are assigned to the original tutor.
    \item \emph{Example}: two WEs on BC are provided.
    \item \emph{Nudge}: same as $Example$, and nudges (shown in Figure \ref{fig:prompt}) are prompted to switch to BC in some problems.
    \item \emph{Presented}: same as $Example$, and students are presented some problems in BC by default.
\end{itemize}

\noindent In Figure \ref{fig:modified_logic}, it is crucial to note that: 1) \emph{white} problems behave the same as the original tutor, 2) \emph{red} problems are selected based on the historical strategy switches in our data, and 3) nudges are prompted after a  number of seconds sampled from a probability distribution of prior students' switch behavior.

\begin{figure}[ht!]
\begin{center}
\includegraphics[width=0.48\textwidth]{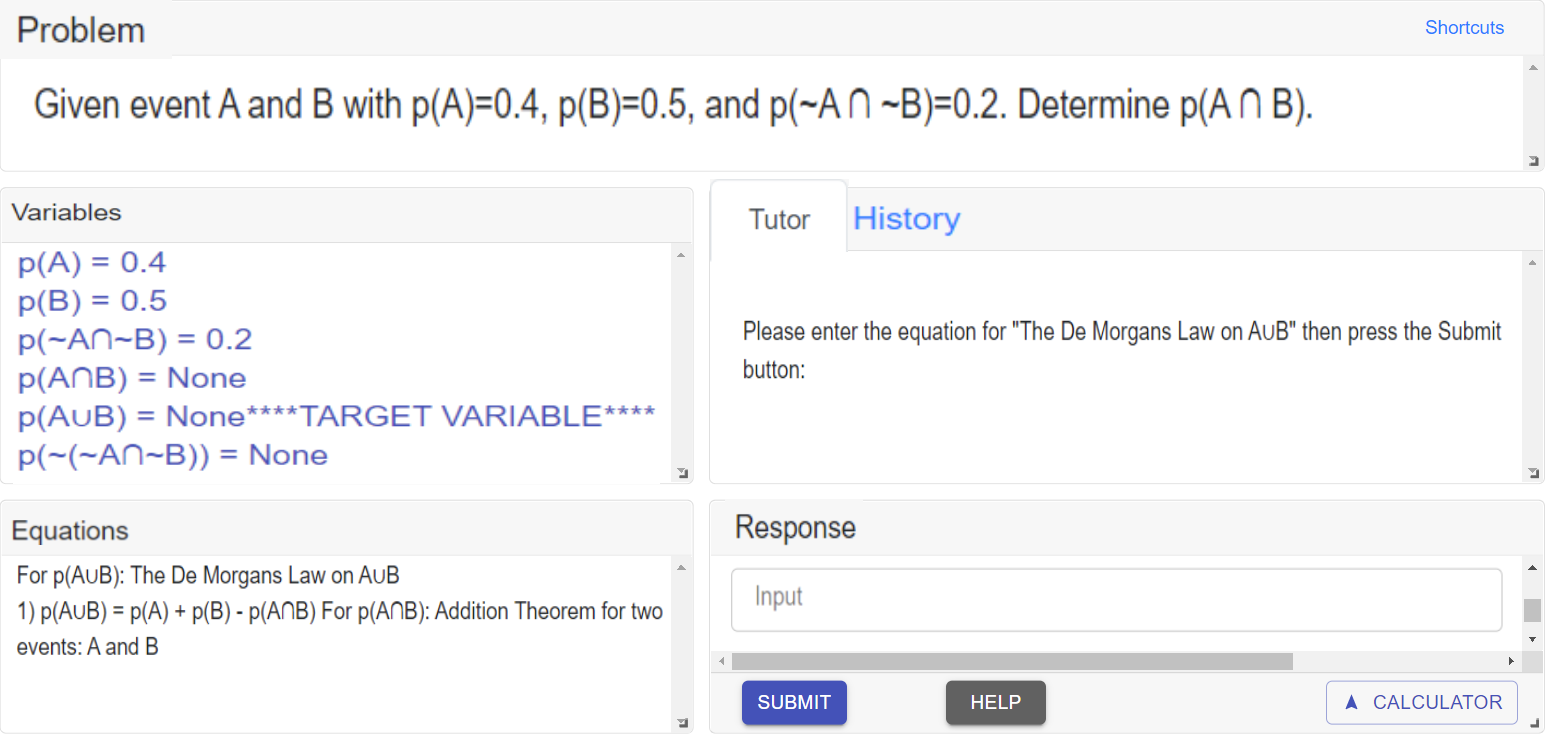}
\end{center}
\caption{Probability Tutor Interface} 
\label{fig:pyr}
\end{figure}

\subsubsection{Probability tutor:} It teaches how to solve probability problems using ten principles, such as the Complement Theorem and De Morgan’s Law, as shown in Figure \ref{fig:pyr}. It consists of $12$ problems, each of which can \emph{only} be solved by BC as it requires deriving an answer by \emph{writing and solving equations} until the target is ultimately reduced to the givens. A problem can be $PS$, $WE$ or $FWE$. $PS$ requires students to solve alone, $WE$ involves a step-by-step solution from the tutor, and $FWE$ demands student and tutor collaboration.

\begingroup
\renewcommand{\arraystretch}{1.7}
\begin{table}[ht!]
\footnotesize
\begin{center} 
\caption{Overview of the Study Procedure} 
\label{fullProcedure} 
\begin{tabular}{l|cc}\hline
  \multirow{7}{*}{Logic} & \multicolumn{2}{c}{Pre-test (2 problems)} \\ 
 \Xcline{2-3}{0.8pt}
 
   & \multicolumn{2}{!{\vrule width 0.5pt}c !{\vrule width 0.5pt}}{\textbf{Training (20 problems)}:} \\
     & \multicolumn{2}{!{\vrule width 0.5pt}c !{\vrule width 0.5pt}}{\cellcolor{gray!60} $Control:$ No Intervention} \\
     & \multicolumn{2}{!{\vrule width 0.5pt}c !{\vrule width 0.5pt}}{\cellcolor{gray!60} $Example$, $Nudge$, $Presented:$ Intervention (Fig. \ref{fig:modified_logic})} \\
     & \multicolumn{2}{!{\vrule width 0.5pt}c !{\vrule width 0.5pt}}{\cellcolor{gray!60} $StrTime:$ No Intervention} \\
  \Cline{0.8pt}{2-3}
  & \multicolumn{2}{c}{Post-test (6 problems, including 2 isomorphic)} \\ \cline{1-3}
  \multicolumn{3}{c}{\textbf{Six weeks later}} \\ \cline{1-3}

  \multirow{4}{*}{Prob.} & \multicolumn{2}{c}{Textbook} \\ \cline{2-3}
  & \multicolumn{2}{c}{Pre-test (14 problems)} \\ \cline{2-3}
     & \multicolumn{2}{!{\vrule width 0.5pt}c !{\vrule width 0.5pt}}{\textbf{Training (12 problems)}:} \\
      & \multicolumn{2}{!{\vrule width 0.5pt}c !{\vrule width 0.5pt}}{On ten problems, students choose $PS/WE/FWE$} \\
  \Cline{0.8pt}{2-3}
  & \multicolumn{2}{c}{Post-test (20 problems, including 14 isomorphic)} \\ \cline{1-3}

\end{tabular}
\end{center} 
\end{table}
\endgroup

\subsection{Procedure}\label{procedureSection}

Table \ref{fullProcedure} summarizes our procedure. During the logic instruction, students went through the standard sequence of pre-test, training and post-test. The first two post-test problems are isomorphic to the two pre-test problems. The \emph{only} difference occurred during training on logic, as shown in Table \ref{fullProcedure}.

Six weeks later, students were trained on the probability tutor following the standard procedure:  textbook, pre-test, training, and post-test. In the textbook, they studied the domain principles; In pre- and post-test, students solved $14$ and $20$  open-ended problems that required them to derive an answer by writing and solving one or more equations. Each pre-test problem has a corresponding isomorphic post-test problem. For the training section, shown in Figure \ref{fig:pyr}, students went through $12$ problems and selected the type on ten of them; two problems were fixed as $PS$. For $FWE$ problems, each step was randomly decided to determine whether the student or tutor should solve it. Note that on both tutors,  the post-test is \emph{much more challenging} than the pre-test, and the problem order is the same for all students.

\subsection{Grading criteria} On logic, a problem score is a function of time, accuracy, and solution length. The \emph{pre-} and \emph{post-test} scores are calculated by averaging the pre- and post-test problem scores. On probability, students' answers are graded by experienced graders in a double-blind manner using a partial-credit rubric, and grades are based \emph{only} on accuracy. The \emph{pre-} and \emph{post-test} scores are the average grades in their respective sections. On both tutors, test scores are in the range of $[0,100]$.

\section{Results}

\subsection{Learning Performance}

\begingroup
\renewcommand{\arraystretch}{1.7}
\begin{table}[ht!]
\scriptsize
\begin{center} 
\caption{Comparing Groups across Tutors} 
\label{performanceSummary} 
\begin{tabular}{ccccc|c} 
\Xhline{4\arrayrulewidth}

& \multicolumn{4}{c|}{Condition} &\\
\cline{2-5}
\makecell{}   & \makecell[t]{$Control$ \\ $(N=17)$} & \makecell[t]{$Example$ \\ $(N=19)$} & \makecell[t]{$Nudge$ \\ $(N=20)$} & \makecell[t]{$Presented$ \\ $(N=15)$} & \makecell[t]{$StrTime$ \\ $(N=40)$}  \\
\hline
\multicolumn{6}{c}{Logic Tutor}\\
\hline
$Pre$ &  $59.1 (19)$ &  $56.9 (25)$ & $60.5(13)$  & $60.4 (15)$ & \cellcolor{gray!30}$60 (18)$ \\
$Iso$-$Post$  & $65.4 (8)$ & $69.7 (7)$  & $\mathbf{89.8 (5)^{*}}$ & $83.4 (4)^{*}$ & \cellcolor{gray!30} $85.3 (6)^{*}$ \\
$Iso$-$NLG$ & $0.04 (.24)$ & $0.09 (.3)$ & $\mathbf{0.4 (.13)^{*}}$ & $0.34 (.14)^{*}$ & \cellcolor{gray!30} $0.35 (.19)^{*}$ \\
$Post$ & $59.9 (9)$ & $65.5 (8)$  & $\mathbf{86.1 (5)^{*}}$ & $80 (5)^{*}$ & \cellcolor{gray!30} $81.7 (6)^{*}$ \\
$NLG$  & -$0.05 (.3)$  & $0.05 (.37)$ & $\mathbf{0.39 (.15)^{*}}$ & $0.29 (.16)^{*}$& \cellcolor{gray!30} $0.3 (.23)^{*}$ \\
$Time$  & $5.5(7)$   & $4.8(4)$ & $5.3(4)$ & $6.2(6)$ & \cellcolor{gray!30} $4.6(7)$  \\

\hline
\multicolumn{6}{c}{Probability Tutor}\\
\hline
$Pre$ &  $79.4 (12)$ & $74.5 (17)$  &  $77 (14)$ & $74.1 (14)$ & \cellcolor{gray!30} $76 (15)$ \\
$Iso$-$Post$ & $73.1 (22)$  & $77 (14)$  & $\mathbf{94.2 (6)^{*}}$  & $85.8 (17)$ & \cellcolor{gray!30} $92.6 (13)^{*}$\\
$Iso$-$NLG$ & -$0.06 (.39)$  & $0.03 (.28)$ & $\mathbf{0.32 (.19)^{*}}$ & $0.16 (.22)$ & \cellcolor{gray!30} $0.28 (.2)^{*}$  \\
$Post$ & $70.3 (20)$  & $73.6 (16)$ & $\mathbf{91.9 (5)^{*}}$  & $83.5 (20)$ & \cellcolor{gray!30} $89.3 (11)^{*}$\\
$NLG$ &  -$0.09 (.36)$ & -$0.04 (.35)$  & $\mathbf{0.27 (.24)^{*}}$ & $0.13 (.23)$ & \cellcolor{gray!30} $0.26 (.17)^{*}$ \\
$Time$ & $4.3(6)$   & $3.9(4)$ & $4.2(5)$ & $3.5(4)$ & \cellcolor{gray!30} $4.4(5)$  \\

\Xhline{4\arrayrulewidth}
\end{tabular} 
\end{center} 
 {\raggedright In a row, bold is for the highest value, and asterisk means significance over no asterisks.\par}
\end{table}
\endgroup

Table \ref{performanceSummary} compares the groups' performance across the two tutors showing the mean and standard deviation of pre- and post-test scores, isomorphic scores, training time in hours, and the learning outcome in terms of the normalized learning gain $(NLG)$ defined as $(NLG = \frac{Post - Pre}{\sqrt{100 - Pre}})$, where 100 is the maximum test score. We refer to pre-test, post-test and NLG scores as $Pre$, $Post$ and $NLG$, respectively. A one-way ANOVA using condition as factor found no significant difference on $Pre$: $\mathit{F}(3,67) = 0.14,\, \mathit{p} = .93$ for logic, and $\mathit{F}(3,67) = 0.49,\, \mathit{p} = .69$ for probability. Similarly, no significant difference was found in the training time on both tutors. In order to measure the students' improvement on isomorphic problems, several repeated measures ANOVA were conducted (one for each group on each tutor) using \{$Pre$, $Iso$-$Post$\} as factor.  Results showed that $Nudge$ and $StrTime$ learned significantly with $\mathit{p} <0.0001$ on both tutors, $Presented$ learned significantly with $\mathit{p} = 0.0001$ on logic and $\mathit{p} =0.02$ on probability. $Example$ and $Control$ did not perform significantly higher on $Iso$-$Post$ than $Pre$ on both tutors. These findings verify the RFC's accuracy, as $StrTime$ learned significantly on both tutors, while $Control$ did not, despite both groups receiving no interventions.

\subsubsection{Comparing Conditions} A comparison between the four conditions in Table \ref{performanceSummary} was essential to assess the performance of $Default$ students. On the logic tutor, a one-way ANCOVA\footnote{General effect size $\eta^2$ was reported for conservative results} using condition as factor and $Pre$ as covariate found a significant difference on $Post$: $\mathit{F}(3,66) = 59.7,\, \mathit{p} < .0001, \, \mathit{\eta}^2 = 0.69$. Follow-up post-hoc analyses with Bonferroni\footnote{Bonferroni was chosen for more conservative results} adjustment\footnote{$(\alpha=.05/10)$ throughout the results section} revealed that $Nudge$ and $Presented$ significantly outperformed $Example$ $(\mathit{t}(37) = 5.9,\, \mathit{p} < .0001$ and $\mathit{t}(32) = 5.2,\, \mathit{p} < .0001)$ as well as $Control$ $(\mathit{t}(35) = 7.8,\, \mathit{p} < .0001$ and $\mathit{t}(30) = 6.3,\, \mathit{p} < .0001)$. No significant difference was found between $Nudge$ and $Presented$, or between $Example$ and $Control$. Similar patterns were observed on $NLG$ using $ANOVA$. These findings show that $Nudge,$ $Presented$ $> Example,$ $Control$.

On the probability tutor, a one-way ANCOVA using condition as factor and $Pre$ as covariate reported a significant difference on $Post$: $\mathit{F}(3,66) = 14.5,\, \mathit{p} < .0001, \, \mathit{\eta}^2 = 0.31$. Subsequent Bonferroni-corrected analyses  showed that $Nudge$ significantly outperformed $Presented$ $(\mathit{t}(33) = 3.6,\, \mathit{p} = .001)$, $Example$ $(\mathit{t}(37) = 5.6,\, \mathit{p} < .0001)$ and $Control$ $(\mathit{t}(35) = 6.2,\, \mathit{p} < .0001)$; meanwhile, $Presented$ significantly surpassed $Example$ and $Control$ $(\mathit{t}(32) = 3.1,\, \mathit{p} = .004$ and $\mathit{t}(30) = 3.4,\, \mathit{p} = .002)$. No significant difference was found between between $Example$ and $Control$. Similar patterns were found using $ANOVA$ on $NLG$. In short, these results show that $Nudge>$ $Presented$ $> Example,$ $Control$.

In essence, $Nudge$ students were the best on both tutors, followed by $Presented$, who learned less on probability. Surprisingly, $Example$ learned no different from $Control$ on both tutors, which signifies the additional instructional support that $Nudge$ and $Presented$ were given on logic.

\subsubsection{Comparing with \textbf{\textit{StrTime}}} To determine whether any condition caught up with $StrTime$ students, post-hoc pairwise analyses were conducted on logic and probability $Post$ using Bonferroni correction. On logic, results revealed that $Nudge$ and $Presented$ caught up with $StrTime$ as no significant difference was found between their $Post$ and that of $StrTime$ $(\mathit{t}(58) = 0.9,\, \mathit{p} = .37$ and $\mathit{t}(53) = 0.3,\, \mathit{p} = .77)$. On the other hand, $StrTime$ significantly outperformed $Example$ $(\mathit{t}(57) = 5.4 ,\, \mathit{p} < .0001)$ and $Control$ $(\mathit{t}(55) = 6.7,\, \mathit{p} < .0001)$. Similar results were found on $NLG$.

On the probability tutor, only $Nudge$ caught up with $StrTime$ as no significant difference was found on $Post$ $(\mathit{t}(58) = 0.2,\, \mathit{p} = .84)$. Meanwhile, $StrTime$ significantly surpassed $Presented$ $(\mathit{t}(53) = 3.1,\, \mathit{p} = .003)$, $Example$ $(\mathit{t}(57) = 5.1,\, \mathit{p} < .0001)$ and $Control$ $(\mathit{t}(55) = 5.7,\, \mathit{p} < .0001)$. Similar patterns were observed on $NLG$.

In brief, $Nudge$ and $Presented$ caught up with $StrTime$ in the presence of our interventions on logic. Only $Nudge$ caught up with $StrTime$ on probability without such interventions. Lastly, $Example$ and $Control$ performed significantly worse than $StrTime$ on both tutors.

\begin{figure}[ht!]
\begin{center}
\includegraphics[width=0.48\textwidth]{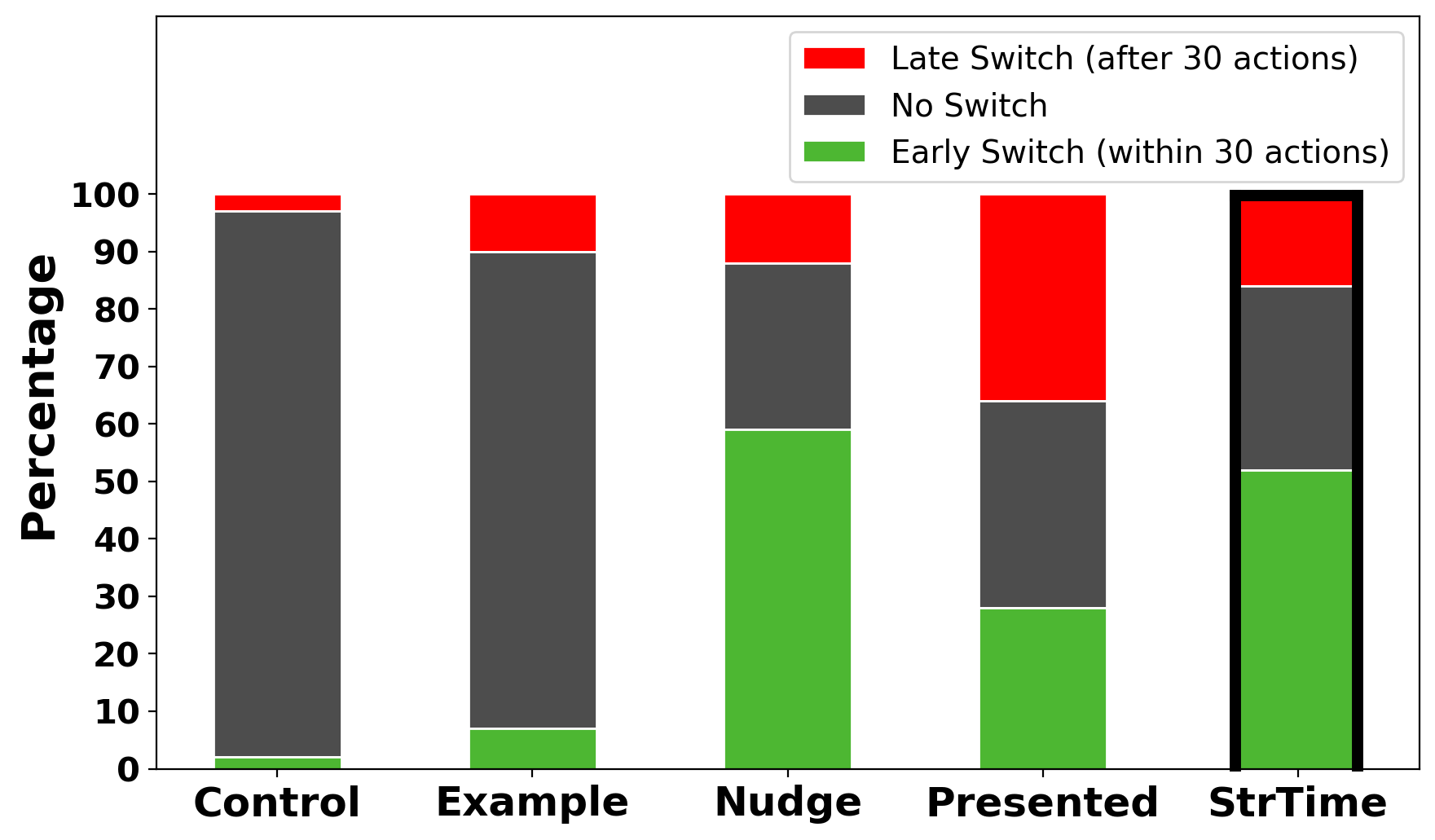}
\end{center}
\caption{Strategy Switch Behavior on Logic} 
\label{fig:switch}
\end{figure}

\subsection{Strategy Switch on Logic}

The strategy switch behavior on the logic tutor (from FC into BC) is displayed in Figure \ref{fig:switch} to investigate the impact of our intervention on students' strategy choices. Decisions are combined across the training and post-test sections, as no significant difference was found in their distribution between the two sections. Additionally, $StrTime$ is highlighted in bold as the gold standard.

A one-way ANOVA using condition as factor showed a significant difference in the frequency of \textit{early} switches: $\mathit{F}(3,67) = 6.7,\, \mathit{p} < .001, \, \mathit{\eta}^2 = 0.23$. Moreover, a chi-square test showed a significant relationship between the switch type and student group\footnote{[$111$ students] * [$20$ training - $2$ WE + $6$ post] = $2664$ decisions}: $\chi^2 (8,\, N=2664) = 934.3, \, \mathit{p}<.0001$. Post-hoc pairwise chi-square tests with Bonferroni adjustment  showed that for \textit{early} switches: $Nudge,$ $StrTime>$ $Presented$ $> Example,$ $Control$. For instance, $Nudge$ and $StrTime$ made early switches significantly more than $Presented$: $\chi^2 (2,\, N=840) = 100.2, \, \mathit{p}<.0001$ and $\chi^2 (2,\, N=1320) = 84.2, \, \mathit{p}<.0001$, respectively. No significant difference was found between $Nudge$ and $StrTime$, or between $Example$ and $Control$.

\begin{figure}[ht!]
\begin{center}
\includegraphics[width=0.48\textwidth]{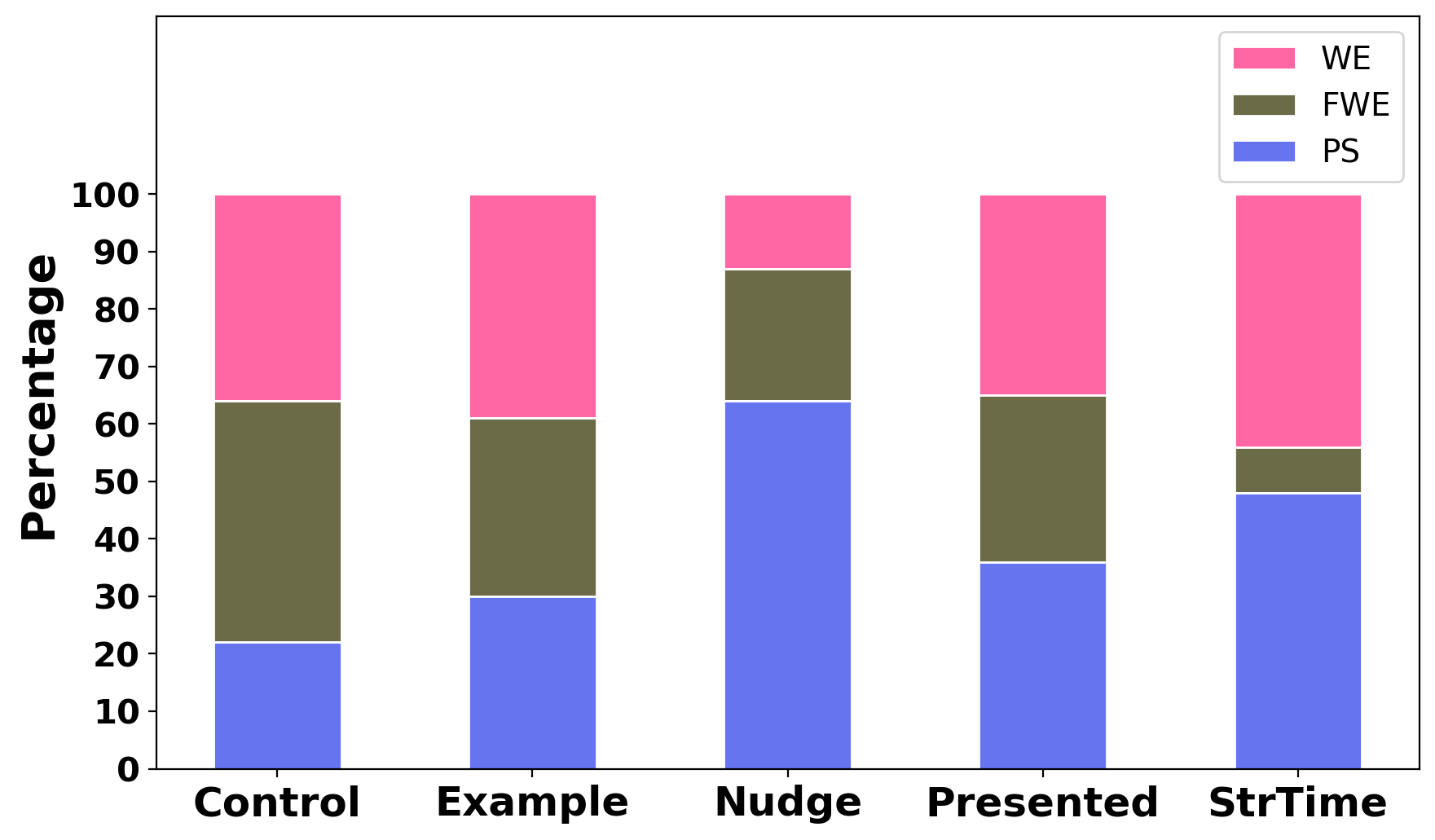}
\end{center}
\caption{Problem-level Decisions on Probability} 
\label{fig:choice}
\end{figure}
\subsection{Student Decision on Probability}

Table \ref{performanceSummary} showed that our interventions' impact on logic might also extend to probability. Therefore, in Figure \ref{fig:choice}, we investigate the problem-level decisions in the probability training section as students chose them. Step-level decisions were not considered since the tutor randomly chose them. It is important to note that for each student group, there was no significant correlation between any problem-level decision type shown in Figure \ref{fig:choice} and any performance metric in Table \ref{performanceSummary}.

A chi-square test found a significant relationship between the problem-level decision type and student group\footnote{[$111$ students] * [$10$ choices on training] = $1110$ decisions}: $\chi^2 (8,\, N=1110) = 162.1, \, \mathit{p}<.0001$. Follow-up pairwise chi-square tests with Bonferroni correction showed that for \textit{PS}: $Nudge>$ $Presented,$ $Example,$ $Control$; for \textit{FWE}: $StrTime<$ $Nudge,$ $Presented,$ $Example,$ $Control$. For instance, $Nudge$ chose $PS$ significantly more than $Presented$: $\chi^2 (2,\, N=350) = 32.9, \, \mathit{p}<.0001$, while $StrTime$ chose $FWE$ significantly less than $Nudge$: $\chi^2 (2,\, N=600) = 67.5, \, \mathit{p}<.0001$. No significant difference was found between any pair of $Presented$, $Example$ and $Control$ on any decision type. In short, $Nudge$ and $StrTime$ made decisions different from each other and their peers, while $Presented$, $Example$ and $Control$ made similar decisions.

\subsection{Discussions \& Conclusions}

We showed that to teach students \emph{how} and \emph{when} to use a strategy, using worked examples alone may not be very effective, as $Example$ did not significantly outperform $Control$. However, students learned better when we reinforced examples by prompting BC nudges or presenting problems in BC by default, as $Nudge$ and $Presented$ significantly surpassed $Example$ and $Control$. Additionally, providing nudges was even more beneficial as $Nudge$ continued to outperform $Presented$ on probability significantly.

\subsubsection{Catching up with StrTime} While $Nudge$ and $Presented$ caught up with $StrTime$ on logic, only $Nudge$ caught up with $StrTime$ on their logic early-switch behavior and probability learning performance. This finding suggests that the $Nudge$ students are prepared for future learning \cite{bransford1999transferRethinking} as they performed well on probability based on interventions they received on logic.

\subsubsection{Relation to ICAP Framework} We believe that our results show that the effectiveness of the Interactive, Constructive, Active and Passive (ICAP) framework \cite{chi2014icap, chi2009ica} can be extended to teaching students metacognitive skills. Precisely, $Control$ encountered passive learning as they received no interventions, while $Example$ received an active treatment as students were required to go through the examples and proceed to the next steps. $Presented$ can be seen as a constructive intervention since students were asked to generate solutions in a strategy presented to them beyond the default one. Finally, $Nudge$ received an interactive intervention where the tutor offered nudges to switch strategies, but the actual switch had to come from students. Our findings are consistent with ICAP in that interactive learning activities achieve the highest learning outcomes, as is the case with $Nudge$ students on the two tutors.

\subsubsection{Relation to Nudge Theory} The nudge theory \cite{thaler2008NudgeTheory} states that nudges have an essential role in behavioral sciences \cite{simon2018NudgeBehavioralSciences} and influence individuals' social and cognitive behavior \cite{smith2013NudgeChoice,goldstein2008nudgeChoice}. Our results suggest that the impact of this theory is evident in teaching $Default$ students the BC strategy on a logic tutor. Precisely, the strategy behavior of $Nudge$ students changed after receiving prompted nudges to use BC, resulting in the best performance on both tutors. 

\subsubsection{Students' Choices and Personalities} The evaluation of students' choices on probability revealed that $StrTime$ students preferred minimal collaboration with the tutor; they chose $WE$ or $PS$ likely to save time or show effort, respectively. On the other hand, $Nudge$ students chose $PS$ significantly more than their intervention and $Control$ peers, likely to demonstrate their acquired BC knowledge. At the end of probability training, students were provided the ten-item personality inventory\footnote{This was not stated earlier for not being our main scope} (TIPI) \cite{gosling2003tipi}, which showed that $Nudge$ and $StrTime$ identified themselves as \textit{critical} and \textit{quarrelsome} significantly more than their peers.

\subsubsection{Limitations and Future Work}

There are at least two caveats in our study. First, our study focused on different interventions for $Default$ students, and hence, the conditions ended up with relatively small sample sizes. Second, the logic tutor offered a strategy by default, and the probability tutor supported only one strategy. A more convincing testbed would be having the tutors support both strategies, where students will be asked to choose the default strategy on each problem. The future work includes combining nudges and presentation into one intervention, implementing FC on the probability tutor, and providing explanations in the nudges on why $BC$ is helpful.

\section{Acknowledgments}
This research was supported by the NSF Grants:
MetaDash: A Teacher Dashboard Informed by Real-Time Multichannel Self-Regulated Learning Data (1660878), Integrated Data-driven Technologies for Individualized Instruction in STEM Learning Environments (1726550), Generalizing Data-Driven Technologies to Improve Individualized STEM Instruction by Intelligent Tutors (2013502) and CAREER: Improving Adaptive Decision Making in Interactive Learning Environments (1651909).

\bibliographystyle{apacite}

\setlength{\bibleftmargin}{.125in}
\setlength{\bibindent}{-\bibleftmargin}

\bibliography{cogsci2022}

\begin{thebibliography}{}

\bibitem [\protect \citeauthoryear {%
Abdelshiheed%
, Maniktala%
, Barnes%
\BCBL {}\ \BBA {} Chi%
}{%
Abdelshiheed%
\ \protect \BOthers {.}}{%
{\protect \APACyear {2022}}%
}]{%
abdelshiheed2022assessing}
\APACinsertmetastar {%
abdelshiheed2022assessing}%
\begin{APACrefauthors}%
Abdelshiheed, M.%
, Maniktala, M.%
, Barnes, T.%
\BCBL {}\ \BBA {} Chi, M.%
\end{APACrefauthors}%
\unskip\
\newblock
\APACrefYearMonthDay{2022}{}{}.
\newblock
{\BBOQ}\APACrefatitle {Assessing Competency Using Metacognition and Motivation:
  The Role of Time-Awareness in Preparation for Future Learning} {Assessing
  competency using metacognition and motivation: The role of time-awareness in
  preparation for future learning}.{\BBCQ}
\newblock
\BIn{} \APACrefbtitle {Design Recommendations for Intelligent Tutoring Systems}
  {Design recommendations for intelligent tutoring systems}\ (\BVOL~9, \BPGS\
  121--131).
\PrintBackRefs{\CurrentBib}

\bibitem [\protect \citeauthoryear {%
Abdelshiheed%
\ \protect \BOthers {.}}{%
Abdelshiheed%
\ \protect \BOthers {.}}{%
{\protect \APACyear {2021}}%
}]{%
abdelshiheed2021preparing}
\APACinsertmetastar {%
abdelshiheed2021preparing}%
\begin{APACrefauthors}%
Abdelshiheed, M.%
, Maniktala, M.%
, Ju, S.%
, Jain, A.%
, Barnes, T.%
\BCBL {}\ \BBA {} Chi, M.%
\end{APACrefauthors}%
\unskip\
\newblock
\APACrefYearMonthDay{2021}{}{}.
\newblock
{\BBOQ}\APACrefatitle {Preparing Unprepared Students For Future Learning}
  {Preparing unprepared students for future learning}.{\BBCQ}
\newblock
\BIn{} \APACrefbtitle {Proceedings of the 43rd annual conference of the
  cognitive science society} {Proceedings of the 43rd annual conference of the
  cognitive science society}\ (\BVOL~43).
\PrintBackRefs{\CurrentBib}

\bibitem [\protect \citeauthoryear {%
Abdelshiheed%
, Zhou%
, Maniktala%
, Barnes%
\BCBL {}\ \BBA {} Chi%
}{%
Abdelshiheed%
\ \protect \BOthers {.}}{%
{\protect \APACyear {2020}}%
}]{%
abdelshiheed2020metacognition}
\APACinsertmetastar {%
abdelshiheed2020metacognition}%
\begin{APACrefauthors}%
Abdelshiheed, M.%
, Zhou, G.%
, Maniktala, M.%
, Barnes, T.%
\BCBL {}\ \BBA {} Chi, M.%
\end{APACrefauthors}%
\unskip\
\newblock
\APACrefYearMonthDay{2020}{}{}.
\newblock
{\BBOQ}\APACrefatitle {Metacognition and Motivation: The Role of Time-Awareness
  in Preparation for Future Learning} {Metacognition and motivation: The role
  of time-awareness in preparation for future learning}.{\BBCQ}
\newblock
\BIn{} \APACrefbtitle {Proceedings of the 42nd annual conference of the
  cognitive science society} {Proceedings of the 42nd annual conference of the
  cognitive science society}\ (\BVOL~42).
\PrintBackRefs{\CurrentBib}

\bibitem [\protect \citeauthoryear {%
Belenky%
\ \BBA {} Nokes%
}{%
Belenky%
\ \BBA {} Nokes%
}{%
{\protect \APACyear {2009}}%
}]{%
belenky2009metacogTransfer}
\APACinsertmetastar {%
belenky2009metacogTransfer}%
\begin{APACrefauthors}%
Belenky, D\BPBI M.%
\BCBT {}\ \BBA {} Nokes, T\BPBI J.%
\end{APACrefauthors}%
\unskip\
\newblock
\APACrefYearMonthDay{2009}{}{}.
\newblock
{\BBOQ}\APACrefatitle {Examining the role of manipulatives and metacognition on
  engagement, learning, and transfer} {Examining the role of manipulatives and
  metacognition on engagement, learning, and transfer}.{\BBCQ}
\newblock
\APACjournalVolNumPages{The Journal of Problem Solving}{2}{2}{6}.
\PrintBackRefs{\CurrentBib}

\bibitem [\protect \citeauthoryear {%
Belenky%
\ \BBA {} Nokes-Malach%
}{%
Belenky%
\ \BBA {} Nokes-Malach%
}{%
{\protect \APACyear {2012}}%
}]{%
belenky2012strategyAwarenessPFL}
\APACinsertmetastar {%
belenky2012strategyAwarenessPFL}%
\begin{APACrefauthors}%
Belenky, D\BPBI M.%
\BCBT {}\ \BBA {} Nokes-Malach, T\BPBI J.%
\end{APACrefauthors}%
\unskip\
\newblock
\APACrefYearMonthDay{2012}{}{}.
\newblock
{\BBOQ}\APACrefatitle {Motivation and transfer: The role of mastery-approach
  goals in preparation for future learning} {Motivation and transfer: The role
  of mastery-approach goals in preparation for future learning}.{\BBCQ}
\newblock
\APACjournalVolNumPages{Journal of the Learning Sciences}{21}{3}{399--432}.
\PrintBackRefs{\CurrentBib}

\bibitem [\protect \citeauthoryear {%
Bransford%
\ \BBA {} Schwartz%
}{%
Bransford%
\ \BBA {} Schwartz%
}{%
{\protect \APACyear {1999}}%
}]{%
bransford1999transferRethinking}
\APACinsertmetastar {%
bransford1999transferRethinking}%
\begin{APACrefauthors}%
Bransford, J\BPBI D.%
\BCBT {}\ \BBA {} Schwartz, D\BPBI L.%
\end{APACrefauthors}%
\unskip\
\newblock
\APACrefYearMonthDay{1999}{}{}.
\newblock
{\BBOQ}\APACrefatitle {Rethinking transfer: A simple proposal with multiple
  implications} {Rethinking transfer: A simple proposal with multiple
  implications}.{\BBCQ}
\newblock
\APACjournalVolNumPages{Review of research in education}{24}{1}{61--100}.
\PrintBackRefs{\CurrentBib}

\bibitem [\protect \citeauthoryear {%
Cardelle-Elawar%
}{%
Cardelle-Elawar%
}{%
{\protect \APACyear {1992}}%
}]{%
cardelle1992effectsMETACOGNITIVE}
\APACinsertmetastar {%
cardelle1992effectsMETACOGNITIVE}%
\begin{APACrefauthors}%
Cardelle-Elawar, M.%
\end{APACrefauthors}%
\unskip\
\newblock
\APACrefYearMonthDay{1992}{}{}.
\newblock
{\BBOQ}\APACrefatitle {Effects of teaching metacognitive skills to students
  with low mathematics ability} {Effects of teaching metacognitive skills to
  students with low mathematics ability}.{\BBCQ}
\newblock
\APACjournalVolNumPages{Teaching and teacher education}{8}{2}{109--121}.
\PrintBackRefs{\CurrentBib}

\bibitem [\protect \citeauthoryear {%
Chambres%
\ \protect \BOthers {.}}{%
Chambres%
\ \protect \BOthers {.}}{%
{\protect \APACyear {2002}}%
}]{%
chambres2002metacognitionStrategySelection}
\APACinsertmetastar {%
chambres2002metacognitionStrategySelection}%
\begin{APACrefauthors}%
Chambres, P.%
\BCBT {}\ \BOthersPeriod {.}
\end{APACrefauthors}%
\unskip\
\newblock
\APACrefYear{2002}.
\newblock
\APACrefbtitle {Metacognition: Process, function, and use} {Metacognition:
  Process, function, and use}.
\newblock
\APACaddressPublisher{}{Kluwer Academic Publishers}.
\PrintBackRefs{\CurrentBib}

\bibitem [\protect \citeauthoryear {%
Chamot%
}{%
Chamot%
}{%
{\protect \APACyear {1998}}%
}]{%
chamot1998strategyAwareness}
\APACinsertmetastar {%
chamot1998strategyAwareness}%
\begin{APACrefauthors}%
Chamot, A.%
\end{APACrefauthors}%
\unskip\
\newblock
\APACrefYearMonthDay{1998}{}{}.
\newblock
{\BBOQ}\APACrefatitle {Teaching learning language strategies to language
  students} {Teaching learning language strategies to language
  students}.{\BBCQ}
\newblock
\APACjournalVolNumPages{Language and Linguistics}{}{}{}.
\PrintBackRefs{\CurrentBib}

\bibitem [\protect \citeauthoryear {%
M.~Chi%
\ \BBA {} VanLehn%
}{%
M.~Chi%
\ \BBA {} VanLehn%
}{%
{\protect \APACyear {2007}}%
}]{%
chi2007pyrenees2}
\APACinsertmetastar {%
chi2007pyrenees2}%
\begin{APACrefauthors}%
Chi, M.%
\BCBT {}\ \BBA {} VanLehn, K.%
\end{APACrefauthors}%
\unskip\
\newblock
\APACrefYearMonthDay{2007}{}{}.
\newblock
{\BBOQ}\APACrefatitle {The impact of explicit strategy instruction on
  problem-solving behaviors across intelligent tutoring systems} {The impact of
  explicit strategy instruction on problem-solving behaviors across intelligent
  tutoring systems}.{\BBCQ}
\newblock
\BIn{} \APACrefbtitle {Proceedings of the Annual Meeting of the Cognitive
  Science Society} {Proceedings of the annual meeting of the cognitive science
  society}\ (\BVOL~29).
\PrintBackRefs{\CurrentBib}

\bibitem [\protect \citeauthoryear {%
M.~Chi%
\ \BBA {} VanLehn%
}{%
M.~Chi%
\ \BBA {} VanLehn%
}{%
{\protect \APACyear {2008}}%
}]{%
chi2008eliminatingGap}
\APACinsertmetastar {%
chi2008eliminatingGap}%
\begin{APACrefauthors}%
Chi, M.%
\BCBT {}\ \BBA {} VanLehn, K.%
\end{APACrefauthors}%
\unskip\
\newblock
\APACrefYearMonthDay{2008}{}{}.
\newblock
{\BBOQ}\APACrefatitle {Eliminating the gap between the high and low students
  through meta-cognitive strategy instruction} {Eliminating the gap between the
  high and low students through meta-cognitive strategy instruction}.{\BBCQ}
\newblock
\BIn{} \APACrefbtitle {Intelligent Tutoring Systems} {Intelligent tutoring
  systems}\ (\BPGS\ 603--613).
\PrintBackRefs{\CurrentBib}

\bibitem [\protect \citeauthoryear {%
M.~Chi%
\ \BBA {} VanLehn%
}{%
M.~Chi%
\ \BBA {} VanLehn%
}{%
{\protect \APACyear {2010}}%
}]{%
chi2010backward2metacogStrategyINSTRUCTION}
\APACinsertmetastar {%
chi2010backward2metacogStrategyINSTRUCTION}%
\begin{APACrefauthors}%
Chi, M.%
\BCBT {}\ \BBA {} VanLehn, K.%
\end{APACrefauthors}%
\unskip\
\newblock
\APACrefYearMonthDay{2010}{}{}.
\newblock
{\BBOQ}\APACrefatitle {Meta-Cognitive Strategy Instruction in Intelligent
  Tutoring Systems: How, When, and Why.} {Meta-cognitive strategy instruction
  in intelligent tutoring systems: How, when, and why.}{\BBCQ}
\newblock
\APACjournalVolNumPages{Educational Technology \& Society}{13}{1}{25--39}.
\PrintBackRefs{\CurrentBib}

\bibitem [\protect \citeauthoryear {%
M\BPBI T.~Chi%
}{%
M\BPBI T.~Chi%
}{%
{\protect \APACyear {2009}}%
}]{%
chi2009ica}
\APACinsertmetastar {%
chi2009ica}%
\begin{APACrefauthors}%
Chi, M\BPBI T.%
\end{APACrefauthors}%
\unskip\
\newblock
\APACrefYearMonthDay{2009}{}{}.
\newblock
{\BBOQ}\APACrefatitle {Active-constructive-interactive: A conceptual framework
  for differentiating learning activities} {Active-constructive-interactive: A
  conceptual framework for differentiating learning activities}.{\BBCQ}
\newblock
\APACjournalVolNumPages{Topics in cognitive science}{1}{1}{73--105}.
\PrintBackRefs{\CurrentBib}

\bibitem [\protect \citeauthoryear {%
M\BPBI T.~Chi%
\ \BBA {} Wylie%
}{%
M\BPBI T.~Chi%
\ \BBA {} Wylie%
}{%
{\protect \APACyear {2014}}%
}]{%
chi2014icap}
\APACinsertmetastar {%
chi2014icap}%
\begin{APACrefauthors}%
Chi, M\BPBI T.%
\BCBT {}\ \BBA {} Wylie, R.%
\end{APACrefauthors}%
\unskip\
\newblock
\APACrefYearMonthDay{2014}{}{}.
\newblock
{\BBOQ}\APACrefatitle {The ICAP framework: Linking cognitive engagement to
  active learning outcomes} {The icap framework: Linking cognitive engagement
  to active learning outcomes}.{\BBCQ}
\newblock
\APACjournalVolNumPages{Educational psychologist}{49}{4}{219--243}.
\PrintBackRefs{\CurrentBib}

\bibitem [\protect \citeauthoryear {%
de Boer%
\ \protect \BOthers {.}}{%
de Boer%
\ \protect \BOthers {.}}{%
{\protect \APACyear {2018}}%
}]{%
de2018longSWITCH-INSTRUCTION}
\APACinsertmetastar {%
de2018longSWITCH-INSTRUCTION}%
\begin{APACrefauthors}%
de Boer, H.%
\BCBT {}\ \BOthersPeriod {.}
\end{APACrefauthors}%
\unskip\
\newblock
\APACrefYearMonthDay{2018}{}{}.
\newblock
{\BBOQ}\APACrefatitle {Long-term effects of metacognitive strategy instruction
  on student academic performance: A meta-analysis} {Long-term effects of
  metacognitive strategy instruction on student academic performance: A
  meta-analysis}.{\BBCQ}
\newblock
\APACjournalVolNumPages{Educational Research Review}{24}{}{98--115}.
\PrintBackRefs{\CurrentBib}

\bibitem [\protect \citeauthoryear {%
Erskine%
}{%
Erskine%
}{%
{\protect \APACyear {2010}}%
}]{%
erskine2010metacognitiveInstruction}
\APACinsertmetastar {%
erskine2010metacognitiveInstruction}%
\begin{APACrefauthors}%
Erskine, D\BPBI L.%
\end{APACrefauthors}%
\unskip\
\newblock
\APACrefYear{2010}.
\newblock
\APACrefbtitle {Effect of prompted reflection and metacognitive skill
  instruction on university freshmen's use of metacognition} {Effect of
  prompted reflection and metacognitive skill instruction on university
  freshmen's use of metacognition}.
\newblock
\APACaddressPublisher{}{Brigham Young University}.
\PrintBackRefs{\CurrentBib}

\bibitem [\protect \citeauthoryear {%
Fazio%
\ \protect \BOthers {.}}{%
Fazio%
\ \protect \BOthers {.}}{%
{\protect \APACyear {2016}}%
}]{%
fazio2016timeAwareness}
\APACinsertmetastar {%
fazio2016timeAwareness}%
\begin{APACrefauthors}%
Fazio, L\BPBI K.%
\BCBT {}\ \BOthersPeriod {.}
\end{APACrefauthors}%
\unskip\
\newblock
\APACrefYearMonthDay{2016}{}{}.
\newblock
{\BBOQ}\APACrefatitle {Strategy use and strategy choice in fraction magnitude
  comparison.} {Strategy use and strategy choice in fraction magnitude
  comparison.}{\BBCQ}
\newblock
\APACjournalVolNumPages{Journal of Experimental Psychology: Learning, Memory,
  and Cognition}{42}{1}{1}.
\PrintBackRefs{\CurrentBib}

\bibitem [\protect \citeauthoryear {%
Fellman%
\ \protect \BOthers {.}}{%
Fellman%
\ \protect \BOthers {.}}{%
{\protect \APACyear {2020}}%
}]{%
fellman2020Presented}
\APACinsertmetastar {%
fellman2020Presented}%
\begin{APACrefauthors}%
Fellman, D.%
, Jylkk{\"a}, J.%
, Waris, O.%
, Soveri, A.%
, Ritakallio, L.%
, Haga, S.%
\BDBL {}Laine, M.%
\end{APACrefauthors}%
\unskip\
\newblock
\APACrefYearMonthDay{2020}{}{}.
\newblock
{\BBOQ}\APACrefatitle {The role of strategy use in working memory training
  outcomes} {The role of strategy use in working memory training
  outcomes}.{\BBCQ}
\newblock
\APACjournalVolNumPages{Journal of Memory and Language}{110}{}{104064}.
\PrintBackRefs{\CurrentBib}

\bibitem [\protect \citeauthoryear {%
Glogger-Frey%
\ \protect \BOthers {.}}{%
Glogger-Frey%
\ \protect \BOthers {.}}{%
{\protect \APACyear {2015}}%
}]{%
glogger2015WE}
\APACinsertmetastar {%
glogger2015WE}%
\begin{APACrefauthors}%
Glogger-Frey, I.%
\BCBT {}\ \BOthersPeriod {.}
\end{APACrefauthors}%
\unskip\
\newblock
\APACrefYearMonthDay{2015}{}{}.
\newblock
{\BBOQ}\APACrefatitle {Inventing a solution and studying a worked solution
  prepare differently for learning from direct instruction} {Inventing a
  solution and studying a worked solution prepare differently for learning from
  direct instruction}.{\BBCQ}
\newblock
\APACjournalVolNumPages{Learning and Instruction}{39}{}{72--87}.
\PrintBackRefs{\CurrentBib}

\bibitem [\protect \citeauthoryear {%
Goldstein%
\ \protect \BOthers {.}}{%
Goldstein%
\ \protect \BOthers {.}}{%
{\protect \APACyear {2008}}%
}]{%
goldstein2008nudgeChoice}
\APACinsertmetastar {%
goldstein2008nudgeChoice}%
\begin{APACrefauthors}%
Goldstein, D\BPBI G.%
\BCBT {}\ \BOthersPeriod {.}
\end{APACrefauthors}%
\unskip\
\newblock
\APACrefYearMonthDay{2008}{}{}.
\newblock
{\BBOQ}\APACrefatitle {Nudge your customers toward better choices} {Nudge your
  customers toward better choices}.{\BBCQ}
\newblock
\APACjournalVolNumPages{Harvard Business Review}{86}{12}{99--105}.
\PrintBackRefs{\CurrentBib}

\bibitem [\protect \citeauthoryear {%
Gosling%
, Rentfrow%
\BCBL {}\ \BBA {} Swann~Jr%
}{%
Gosling%
\ \protect \BOthers {.}}{%
{\protect \APACyear {2003}}%
}]{%
gosling2003tipi}
\APACinsertmetastar {%
gosling2003tipi}%
\begin{APACrefauthors}%
Gosling, S\BPBI D.%
, Rentfrow, P\BPBI J.%
\BCBL {}\ \BBA {} Swann~Jr, W\BPBI B.%
\end{APACrefauthors}%
\unskip\
\newblock
\APACrefYearMonthDay{2003}{}{}.
\newblock
{\BBOQ}\APACrefatitle {A very brief measure of the Big-Five personality
  domains} {A very brief measure of the big-five personality domains}.{\BBCQ}
\newblock
\APACjournalVolNumPages{Journal of Research in personality}{37}{6}{504--528}.
\PrintBackRefs{\CurrentBib}

\bibitem [\protect \citeauthoryear {%
Lee%
\ \BBA {} Oxford%
}{%
Lee%
\ \BBA {} Oxford%
}{%
{\protect \APACyear {2008}}%
}]{%
lee2008koreanStrategyAwareness}
\APACinsertmetastar {%
lee2008koreanStrategyAwareness}%
\begin{APACrefauthors}%
Lee, K\BPBI R.%
\BCBT {}\ \BBA {} Oxford, R.%
\end{APACrefauthors}%
\unskip\
\newblock
\APACrefYearMonthDay{2008}{}{}.
\newblock
{\BBOQ}\APACrefatitle {Understanding EFL Learners’ Strategy Use and Strategy
  Awareness} {Understanding efl learners’ strategy use and strategy
  awareness}.{\BBCQ}
\newblock
\APACjournalVolNumPages{The Asian EFL Journal Quarterly March 2008 Volume 10,
  Issue}{10}{1}{7--32}.
\PrintBackRefs{\CurrentBib}

\bibitem [\protect \citeauthoryear {%
Likourezos%
\ \BBA {} Kalyuga%
}{%
Likourezos%
\ \BBA {} Kalyuga%
}{%
{\protect \APACyear {2017}}%
}]{%
likourezos2017WE}
\APACinsertmetastar {%
likourezos2017WE}%
\begin{APACrefauthors}%
Likourezos, V.%
\BCBT {}\ \BBA {} Kalyuga, S.%
\end{APACrefauthors}%
\unskip\
\newblock
\APACrefYearMonthDay{2017}{}{}.
\newblock
{\BBOQ}\APACrefatitle {Instruction-first and problem-solving-first approaches:
  alternative pathways to learning complex tasks} {Instruction-first and
  problem-solving-first approaches: alternative pathways to learning complex
  tasks}.{\BBCQ}
\newblock
\APACjournalVolNumPages{Instructional Science}{45}{}{195--219}.
\PrintBackRefs{\CurrentBib}

\bibitem [\protect \citeauthoryear {%
Priest%
\ \BBA {} Lindsay%
}{%
Priest%
\ \BBA {} Lindsay%
}{%
{\protect \APACyear {1992}}%
}]{%
priest1992newINSTRUCTION}
\APACinsertmetastar {%
priest1992newINSTRUCTION}%
\begin{APACrefauthors}%
Priest, A.%
\BCBT {}\ \BBA {} Lindsay, R.%
\end{APACrefauthors}%
\unskip\
\newblock
\APACrefYearMonthDay{1992}{}{}.
\newblock
{\BBOQ}\APACrefatitle {New light on novice—expert differences in physics
  problem solving} {New light on novice—expert differences in physics problem
  solving}.{\BBCQ}
\newblock
\APACjournalVolNumPages{British journal of Psychology}{83}{3}{389--405}.
\PrintBackRefs{\CurrentBib}

\bibitem [\protect \citeauthoryear {%
Richey%
\ \protect \BOthers {.}}{%
Richey%
\ \protect \BOthers {.}}{%
{\protect \APACyear {2015}}%
}]{%
richey2015promptsWE}
\APACinsertmetastar {%
richey2015promptsWE}%
\begin{APACrefauthors}%
Richey, J\BPBI E.%
\BCBT {}\ \BOthersPeriod {.}
\end{APACrefauthors}%
\unskip\
\newblock
\APACrefYearMonthDay{2015}{}{}.
\newblock
{\BBOQ}\APACrefatitle {Transfer effects of prompted and self-reported
  analogical comparison and self-explanation.} {Transfer effects of prompted
  and self-reported analogical comparison and self-explanation.}{\BBCQ}
\newblock
\BIn{} \APACrefbtitle {Proceedings of the Annual Meeting of the Cognitive
  Science Society} {Proceedings of the annual meeting of the cognitive science
  society}\ (\BVOL~37).
\PrintBackRefs{\CurrentBib}

\bibitem [\protect \citeauthoryear {%
Roberts%
\ \BBA {} Erdos%
}{%
Roberts%
\ \BBA {} Erdos%
}{%
{\protect \APACyear {1993}}%
}]{%
roberts1993metacogDefinitionStrategySelection2}
\APACinsertmetastar {%
roberts1993metacogDefinitionStrategySelection2}%
\begin{APACrefauthors}%
Roberts, M\BPBI J.%
\BCBT {}\ \BBA {} Erdos, G.%
\end{APACrefauthors}%
\unskip\
\newblock
\APACrefYearMonthDay{1993}{}{}.
\newblock
{\BBOQ}\APACrefatitle {Strategy selection and metacognition} {Strategy
  selection and metacognition}.{\BBCQ}
\newblock
\APACjournalVolNumPages{Educational Psychology}{13}{}{259--266}.
\PrintBackRefs{\CurrentBib}

\bibitem [\protect \citeauthoryear {%
Russell%
\ \BBA {} Norvig%
}{%
Russell%
\ \BBA {} Norvig%
}{%
{\protect \APACyear {2020}}%
}]{%
russell2020artificial}
\APACinsertmetastar {%
russell2020artificial}%
\begin{APACrefauthors}%
Russell, S\BPBI J.%
\BCBT {}\ \BBA {} Norvig, P.%
\end{APACrefauthors}%
\unskip\
\newblock
\APACrefYear{2020}.
\newblock
\APACrefbtitle {Artificial Intelligence: a modern approach} {Artificial
  intelligence: a modern approach}\ (\PrintOrdinal{4}\ \BEd).
\newblock
\APACaddressPublisher{}{Pearson}.
\PrintBackRefs{\CurrentBib}

\bibitem [\protect \citeauthoryear {%
Schraw%
\ \BBA {} Gutierrez%
}{%
Schraw%
\ \BBA {} Gutierrez%
}{%
{\protect \APACyear {2015}}%
}]{%
schraw2015metacognitiveInstruction}
\APACinsertmetastar {%
schraw2015metacognitiveInstruction}%
\begin{APACrefauthors}%
Schraw, G.%
\BCBT {}\ \BBA {} Gutierrez, A\BPBI P.%
\end{APACrefauthors}%
\unskip\
\newblock
\APACrefYearMonthDay{2015}{}{}.
\newblock
{\BBOQ}\APACrefatitle {Metacognitive strategy instruction that highlights the
  role of monitoring and control processes} {Metacognitive strategy instruction
  that highlights the role of monitoring and control processes}.{\BBCQ}
\newblock
\BIn{} \APACrefbtitle {Metacognition: Fundaments, applications, and trends}
  {Metacognition: Fundaments, applications, and trends}\ (\BPGS\ 3--16).
\newblock
\APACaddressPublisher{}{Springer}.
\PrintBackRefs{\CurrentBib}

\bibitem [\protect \citeauthoryear {%
Schwartz%
\ \BBA {} Martin%
}{%
Schwartz%
\ \BBA {} Martin%
}{%
{\protect \APACyear {2004}}%
}]{%
schwartz2004Presented}
\APACinsertmetastar {%
schwartz2004Presented}%
\begin{APACrefauthors}%
Schwartz, D\BPBI L.%
\BCBT {}\ \BBA {} Martin, T.%
\end{APACrefauthors}%
\unskip\
\newblock
\APACrefYearMonthDay{2004}{}{}.
\newblock
{\BBOQ}\APACrefatitle {Inventing to prepare for future learning: The hidden
  efficiency of encouraging original student production in statistics
  instruction} {Inventing to prepare for future learning: The hidden efficiency
  of encouraging original student production in statistics instruction}.{\BBCQ}
\newblock
\APACjournalVolNumPages{Cognition and instruction}{22}{2}{129--184}.
\PrintBackRefs{\CurrentBib}

\bibitem [\protect \citeauthoryear {%
Simon%
\ \BBA {} Tagliabue%
}{%
Simon%
\ \BBA {} Tagliabue%
}{%
{\protect \APACyear {2018}}%
}]{%
simon2018NudgeBehavioralSciences}
\APACinsertmetastar {%
simon2018NudgeBehavioralSciences}%
\begin{APACrefauthors}%
Simon, C.%
\BCBT {}\ \BBA {} Tagliabue, M.%
\end{APACrefauthors}%
\unskip\
\newblock
\APACrefYearMonthDay{2018}{}{}.
\newblock
{\BBOQ}\APACrefatitle {Feeding the behavioral revolution: Contributions of
  behavior analysis to nudging and vice versa} {Feeding the behavioral
  revolution: Contributions of behavior analysis to nudging and vice
  versa}.{\BBCQ}
\newblock
\APACjournalVolNumPages{Journal of Behavioral Economics for
  Policy}{2}{1}{91--97}.
\PrintBackRefs{\CurrentBib}

\bibitem [\protect \citeauthoryear {%
Smith%
\ \protect \BOthers {.}}{%
Smith%
\ \protect \BOthers {.}}{%
{\protect \APACyear {2013}}%
}]{%
smith2013NudgeChoice}
\APACinsertmetastar {%
smith2013NudgeChoice}%
\begin{APACrefauthors}%
Smith, N\BPBI C.%
\BCBT {}\ \BOthersPeriod {.}
\end{APACrefauthors}%
\unskip\
\newblock
\APACrefYearMonthDay{2013}{}{}.
\newblock
{\BBOQ}\APACrefatitle {Choice without awareness: Ethical and policy
  implications of defaults} {Choice without awareness: Ethical and policy
  implications of defaults}.{\BBCQ}
\newblock
\APACjournalVolNumPages{Journal of Public Policy \&
  Marketing}{32}{2}{159--172}.
\PrintBackRefs{\CurrentBib}

\bibitem [\protect \citeauthoryear {%
Sp{\"o}rer%
\ \protect \BOthers {.}}{%
Sp{\"o}rer%
\ \protect \BOthers {.}}{%
{\protect \APACyear {2009}}%
}]{%
sporer2009Presented}
\APACinsertmetastar {%
sporer2009Presented}%
\begin{APACrefauthors}%
Sp{\"o}rer, N.%
\BCBT {}\ \BOthersPeriod {.}
\end{APACrefauthors}%
\unskip\
\newblock
\APACrefYearMonthDay{2009}{}{}.
\newblock
{\BBOQ}\APACrefatitle {Improving students' reading comprehension skills:
  Effects of strategy instruction and reciprocal teaching} {Improving students'
  reading comprehension skills: Effects of strategy instruction and reciprocal
  teaching}.{\BBCQ}
\newblock
\APACjournalVolNumPages{Learning and instruction}{19}{3}{272--286}.
\PrintBackRefs{\CurrentBib}

\bibitem [\protect \citeauthoryear {%
Teng%
}{%
Teng%
}{%
{\protect \APACyear {2020}}%
}]{%
teng2020strategyAwareness}
\APACinsertmetastar {%
teng2020strategyAwareness}%
\begin{APACrefauthors}%
Teng, F.%
\end{APACrefauthors}%
\unskip\
\newblock
\APACrefYearMonthDay{2020}{}{}.
\newblock
{\BBOQ}\APACrefatitle {The benefits of metacognitive reading strategy awareness
  instruction for young learners of English as a second language} {The benefits
  of metacognitive reading strategy awareness instruction for young learners of
  english as a second language}.{\BBCQ}
\newblock
\APACjournalVolNumPages{Literacy}{54}{1}{29--39}.
\PrintBackRefs{\CurrentBib}

\bibitem [\protect \citeauthoryear {%
Thaler%
\ \BBA {} Sunstein%
}{%
Thaler%
\ \BBA {} Sunstein%
}{%
{\protect \APACyear {2008}}%
}]{%
thaler2008NudgeTheory}
\APACinsertmetastar {%
thaler2008NudgeTheory}%
\begin{APACrefauthors}%
Thaler, R\BPBI H.%
\BCBT {}\ \BBA {} Sunstein, C\BPBI R.%
\end{APACrefauthors}%
\unskip\
\newblock
\APACrefYearMonthDay{2008}{}{}.
\newblock
\APACrefbtitle {Nudge: Improving decisions about health, wealth, and
  happiness.} {Nudge: Improving decisions about health, wealth, and happiness.}
\newblock
\APACaddressPublisher{}{HeinOnline}.
\PrintBackRefs{\CurrentBib}

\bibitem [\protect \citeauthoryear {%
Vanlehn%
}{%
Vanlehn%
}{%
{\protect \APACyear {2006}}%
}]{%
vanlehn2006behavior}
\APACinsertmetastar {%
vanlehn2006behavior}%
\begin{APACrefauthors}%
Vanlehn, K.%
\end{APACrefauthors}%
\unskip\
\newblock
\APACrefYearMonthDay{2006}{}{}.
\newblock
{\BBOQ}\APACrefatitle {The behavior of tutoring systems} {The behavior of
  tutoring systems}.{\BBCQ}
\newblock
\APACjournalVolNumPages{International journal of artificial intelligence in
  education}{16}{3}{227--265}.
\PrintBackRefs{\CurrentBib}

\bibitem [\protect \citeauthoryear {%
Winne%
\ \BBA {} Azevedo%
}{%
Winne%
\ \BBA {} Azevedo%
}{%
{\protect \APACyear {2014}}%
}]{%
winne2014switchMetacognitiveSWITCH}
\APACinsertmetastar {%
winne2014switchMetacognitiveSWITCH}%
\begin{APACrefauthors}%
Winne, P\BPBI H.%
\BCBT {}\ \BBA {} Azevedo, R.%
\end{APACrefauthors}%
\unskip\
\newblock
\APACrefYearMonthDay{2014}{}{}.
\newblock
{\BBOQ}\APACrefatitle {Metacognition} {Metacognition}.{\BBCQ}
\newblock
\BIn{} R\BPBI K.~Sawyer\ (\BED), \APACrefbtitle {The Cambridge Handbook of the
  Learning Sciences} {The cambridge handbook of the learning sciences}\
  (\PrintOrdinal{2}\ \BEd).
\newblock
\APACaddressPublisher{}{Cambridge University Press}.
\PrintBackRefs{\CurrentBib}

\bibitem [\protect \citeauthoryear {%
Zepeda%
\ \protect \BOthers {.}}{%
Zepeda%
\ \protect \BOthers {.}}{%
{\protect \APACyear {2015}}%
}]{%
zepeda2015INSTRUCTION}
\APACinsertmetastar {%
zepeda2015INSTRUCTION}%
\begin{APACrefauthors}%
Zepeda, C\BPBI D.%
\BCBT {}\ \BOthersPeriod {.}
\end{APACrefauthors}%
\unskip\
\newblock
\APACrefYearMonthDay{2015}{}{}.
\newblock
{\BBOQ}\APACrefatitle {Direct instruction of metacognition benefits adolescent
  science learning, transfer, and motivation: An in vivo study.} {Direct
  instruction of metacognition benefits adolescent science learning, transfer,
  and motivation: An in vivo study.}{\BBCQ}
\newblock
\APACjournalVolNumPages{Journal of Educational Psychology}{107}{4}{954}.
\PrintBackRefs{\CurrentBib}

\bibitem [\protect \citeauthoryear {%
Zimmerman%
}{%
Zimmerman%
}{%
{\protect \APACyear {1990}}%
}]{%
zimmerman1990Metacognition}
\APACinsertmetastar {%
zimmerman1990Metacognition}%
\begin{APACrefauthors}%
Zimmerman, B\BPBI J.%
\end{APACrefauthors}%
\unskip\
\newblock
\APACrefYearMonthDay{1990}{}{}.
\newblock
{\BBOQ}\APACrefatitle {Self-regulated learning and academic achievement: An
  overview} {Self-regulated learning and academic achievement: An
  overview}.{\BBCQ}
\newblock
\APACjournalVolNumPages{Educational psychologist}{25}{1}{3--17}.
\PrintBackRefs{\CurrentBib}

\end{thebibliography}

\end{document}